\documentclass{aa}  

\usepackage{graphicx}
\usepackage{txfonts}
\usepackage[]{hyperref}
\usepackage[]{xcolor}
\usepackage{url}
\begin{document} 

    \title{\emph{RadioAstron} reveals a spine-sheath jet structure in 3C\,273}

   \author{G. Bruni\inst{1}\thanks{Contact e-mail:\href{mailto:gabriele.bruni@inaf.it}{gabriele.bruni@inaf.it}}
          \and
          J. L. G\'omez\inst{2}
          \and
          L. Vega-Garc\'ia\inst{3}
          \and 
          A.~P. Lobanov\inst{3,7}
          \and
          A. Fuentes\inst{2}
          \and 
          T. Savolainen\inst{4,5,3}
          \and
          Y.~Y. Kovalev\inst{6,7,3}
          \and
          M. Perucho\inst{8,9}
          \and
          J.-M. Mart\'i\inst{8,9}
          \and 
          J.~M. Anderson\inst{10,11}
          \and
          P.~G. Edwards\inst{12}
          \and
          L.~I. Gurvits\inst{13,14,12}
          \and
          M.~M. Lisakov\inst{3,6}
          \and
          A.~B. Pushkarev\inst{15,6}
          \and
          K.~V. Sokolovsky\inst{16,17}
          \and
          J.~A. Zensus\inst{3}
          }

   \institute{INAF - Istituto di Astrofisica e Planetologia Spaziali, via del Fosso del Cavaliere, 100, 00133, Rome, Italy\\
   \email{gabriele.bruni@inaf.it}
   \and IAA - Instituto de Astrof\'isica de Andaluc\'ia-CSIC, Glorieta de la Astronom\'ia s/n, E-18008 Granada, Spain
   \and MPIfR - Max Planck Institute for Radio Astronomy, auf dem H\"ugel, 69, 53121, Bonn, Germany
   \and Aalto University Mets\"ahovi Radio Observatory, Mets\"ahovintie 114, 02540 Kylm\"al\"a, Finland
   \and Aalto University Department of Electronics and Nanoengineering, PL 15500, 00076 Aalto, Finland
   \and 
   Lebedev Physical Institute of the Russian Academy of Sciences,
Leninsky prospekt 53, 119991 Moscow, Russia
   \and Moscow Institute of Physics and Technology, Institutsky per.~9, Dolgoprudny, 141700 Moscow region, Russia
   \and Departament d'Astronomia i Astrof\'isica, Universitat de València, C/ Dr. Moliner, 50, 46100 Burjassot, València, Spain
   \and Observatori Astron\`omic, Universitat de València, C/ Catedràtic Jos\'e Beltran, 2, 46980 Paterna, València, Spain
   \and Technische Universität Berlin, Institut für Geodäsie und Geoinformationstechnik, Fakultät VI, Sekr. KAI 2-2, Kaiserin-Augusta-Allee 104-106, D-10553 Berlin, Germany
   \and GFZ German Research Centre for Geosciences, Telegrafenberg, D-14473 Potsdam, Germany
   \and CSIRO Astronomy and Space Science, PO Box 76, Epping, NSW 1710, Australia
   \and JIVE - Joint Institute for VLBI ERIC, Oude Hoogeveensedijk 4, 7991 PD Dwingekoo, The Netherlands
   \and Dept. of Astrodynamics and Space Missions, Delft University of Technology, Kluyverweg 1, 2629 HS Delft, The Netherlands
  \and Crimean Astrophysical Observatory, Nauchny 298409, Crimea, Russia
   \and Department of Physics and Astronomy, Michigan State University, East Lansing,
Michigan 48824, USA
   \and Sternberg Astronomical Institute, Moscow State University, Universitetskii pr. 13, 119992 Moscow, Russia
      }

   \date{Received 14 September 2020; accepted \dots}

%%%%%%%%%%%%%%%%%%%%%%%%%%%%%%%%%%%%%%%%%%%%%%%%%%%%%%%%%%%%%%%%%%%%%%%%%%%%%%%%%%%%%%%%
% \abstract{}{}{}{}{} 
% 5 {} token are mandatory
 
  \abstract{
   We present Space-VLBI \emph{RadioAstron} observations at 1.6 GHz and 4.8 GHz of the flat spectrum radio quasar 3C\,273, with detections on baselines up to 4.5 and 3.3 Earth Diameters, respectively. Achieving the best angular resolution at 1.6 GHz to date, we have imaged limb-brightening in the jet, not previously detected in this source. In contrast, at 4.8 GHz, we detected emission from a central stream of plasma, with a spatial distribution complementary to the limb-brightened emission, indicating an origin in the spine of the jet. While a stratification across the jet width in the flow density, internal energy, magnetic field, or bulk flow velocity are usually invoked to explain the limb-brightening, the different jet structure detected at the two frequencies probably requires a stratification in the emitting electron energy distribution. Future dedicated numerical simulations will allow the determination of which combination of physical parameters are needed to reproduce the spine/sheath structure observed by Space-VLBI with \emph{RadioAstron} in 3C\,273.
   }
   \maketitle

%%%%%%%%%%%%%%%%%%%%%%%%%%%%%%%%%%%%%%%%%%%%%%%%%%%%%%%%%%%%%%%%%%%%%%%%%%%%%%%%%%%%%
%%%%%%%%%%%%%%%%%%%%%%%%%%%%%%%%%%%%%%%%%%%%%%%%%%%%%%%%%%%%%%%%%%%%%%%%%%%%%%%%%%%%%

\section{Introduction}

About 10\% of Active Galactic Nuclei (AGN) have prominent relativistic jets of plasma extending up to megaparsec-scale distances from the supermassive black hole (SMBH) which powers them (see e.g. \citealt{2007ApJ...658..815S} and references therein). These are readily studied at radio wavelengths, and propagate in the inner near-nucleus segment as an outflow with a parabolic shape switching to a nearly conical one down the stream  \citep[e.g.][]{Asada12,2020MNRAS.495.3576K,2020MNRAS.498.2532N}. The launching, acceleration, and collimation of such relativistic jets are the subjects of ongoing theoretical and observational studies. The two most accredited models are the \cite{1977MNRAS.179..433B} mechanism, in which the rotational energy of the black hole powers jet launching, and the \cite{1982MNRAS.199..883B} mechanism, in which the accretion disk produces a magnetically driven wind. More recently, a spine-sheath model, involving a stratified inner jet, has been proposed and developed by different authors (e.g. \citealt{1989A&A...224...24P,1989MNRAS.237..411S,2001MNRAS.321L...1C,2005A&A...432..401G,2008MNRAS.386..945T,2009ApJ...697..985D,2012RAA....12..817X,2015MNRAS.450.2824M}). In this scenario, the jet flow consists of two different fluids: a fast, low-density component, streaming along the central axis (spine) and emerging from the immediate vicinity of the black hole, and a slower, denser component at the edges (sheath) of the conical jet, emerging from the accretion disk. This would imply that both the \citet{1977MNRAS.179..433B} and the \citet{1982MNRAS.199..883B} mechanisms can be invoked to produce the two components of the outflow \citep{2007ApJ...664...26H,2012RAA....12..817X}. Notably, as an effect of relativistic Doppler boosting on photons, the spine would dominate the overall jet emission in blazars (having a small viewing angles from the jet axis), while the sheath would be more evident in radio galaxies (seen at large viewing angles).  

The Very Long Baseline Interferometry (VLBI) technique enables the parsec-scale structure of nearby AGN jets to be resolved and the structure of such outflows to be studied in great detail, including the inner regions in the proximity of the jet base. This powerful technique has allowed prominent emission from the jet edges --- also known as limb-brightening, and related to the sheath structure mentioned above --- to be detected in a number of sources. Among the pioneering works on this topic, \cite{1999ApJ...518L..87A} presented the first linear polarization observations of a spine/sheath structure in the quasar 1055+018, performed with the Very Long Baseline Array (VLBA). Later, a similar structure in the BL Lac Mrk\,501 was found by different authors: \cite{2004ApJ...600..127G} through observations with the first dedicated Space VLBI (SVLBI) mission, VSOP/HALCA, revealed a limb brightening in the jet structure, and explained that in terms of a velocity gradient in the jet. Moreover, \cite{2005MNRAS.356..859P} confirmed the result detecting a spine-sheath polarization structure with the VLBA. More recently, \cite{2016A&A...588L...9B} presented Global Millimeter VLBI Array (GMVA) observations of Cygnus\,A at 86\,GHz, revealing a limb-brightening in the jet flow, with a transverse width suggesting a launching point in the accretion disk rather than in the SMBH vicinity. \cite{2018A&A...616A.188K} stacked five GMVA epochs to image the jet base in M87, finding that the limb-brightened structure could be anchored in the inner portion of the accretion disk, similarly to Cygnus\,A. A stratification of the jet flow in the same source was previously found by \cite{2016A&A...595A..54M}, through the kinematic analysis of multiple VLBA images at 43\,GHz. 
Finally, \citet{2018NatAs...2..472G} used Space VLBI \emph{RadioAstron} observations (see below) to reveal a bright outer jet layer in 3C\,84, with a wide jet base suggesting either a rapid lateral expansion of the jet within 100\,$r_\mathrm{g}$ from the black hole or an origin in the accretion disk.

The \emph{RadioAstron} (hereafter RA) Space VLBI mission \citep{2013ARep...57..153K}, led by the Astro Space Center (ASC, Moscow, Russia) and the Lavochkin Scientific and Production Association (Khimki, Russia), operated between 2011 and 2019. With a diameter of 10 meters, the RA space radio telescope performed interferometric observations with arrays of ground radio telescopes, with a maximum Earth-space baseline at apogee of $\sim$350,000 km. It operated at 0.32\,GHz (P-band), 1.6\,GHz (L-band), 4.8\,GHz (C-band), and 22\,GHz (K-band). Three Key Science Programmes (KSPs) on AGN imaging have collected data since 2013 to study the launching, collimation, and magnetic field properties of jets in known AGN (see \citealt{2020AdSpR..65..712B} for a summary of previous results and observed targets) while the AGN survey studied properties of their cores \citep[e.g.][]{2016ApJ...820L...9K,2020AdSpR..65..705K}. In particular, the RA AGN polarization KSP aims to probe the jet structure and magnetic fields configuration at angular resolution down to a few tens of $\mu$as, in AGN known to have the most prominent polarization properties. More than 20 imaging experiments have been performed. Within the project, observations at 22 GHz of BL Lac from the first observing period (Announcement of Opportunity 1, AO-1, July 2013 -- June 2014) were presented in \citet{2016ApJ...817...96G}, producing the image with the highest angular resolution to date (21\,$\mu$as) and revealing helical magnetic fields in the jet. 

The flat-spectrum radio source 3C\,273, the subject of this work, is the first identified quasar \citep{1963Natur.197.1037H,1963Natur.197.1040O,1963Natur.197.1040S}. One of the most observed VLBI targets, 3C\,273 offered an opportunity for studying the jet cross-section morphology with VSOP/HALCA \citep{2001Sci...294..128L}. The quasar 3C\,273 was also investigated by RA in a non-imaging mode. These observations at 18, 6 and 1.3~cm resulted in the detection of the brightness temperatures in the core of the source exceeding the Inverse Compton limit \citep{2016ApJ...820L...9K} and potential refractive substructure \citep{2016ApJ...820L..10J}. The source was targeted twice as a part of the RA AGN polarization KSP: \citet{2017A&A...604A.111B} performed 22\,GHz observations of 3C\,273, showing a brightness temperature drop of two orders of magnitude in only one year compared to \citet{2016ApJ...820L...9K}. 

Here we present a two-frequency, 1.6 and 4.8\,GHz, study of the jet structure in 3C\,273. Taking advantage of the unprecedented angular resolution offered by RA at 1.6 GHz, we imaged for the first time a limb-brightened jet structure for this source, with a spatial distribution complementary to a spine-dominated emission detected at 4.8\,GHz during previous RA observations, performed in the framework of the Strong AGN KSP. A comparison with the 4.8~GHz RA images allows us to consider a new insight in the properties of the jet flow.

%%%%%%%%%%%%%%%%%%%%%%%%%%%%%%%%%%%%%%%%%%%%%%%%%%%%%%%%%%%%%%%%%%%%%%%%%%%%%%%%%%%%%%%%
%%%%%%%%%%%%%%%%%%%%%%%%%%%%%%%%%%%%%%%%%%%%%%%%%%%%%%%%%%%%%%%%%%%%%%%%%%%%%%%%%%%%%%%%

\begin{figure*}
\centering
\includegraphics[width=0.48\textwidth]{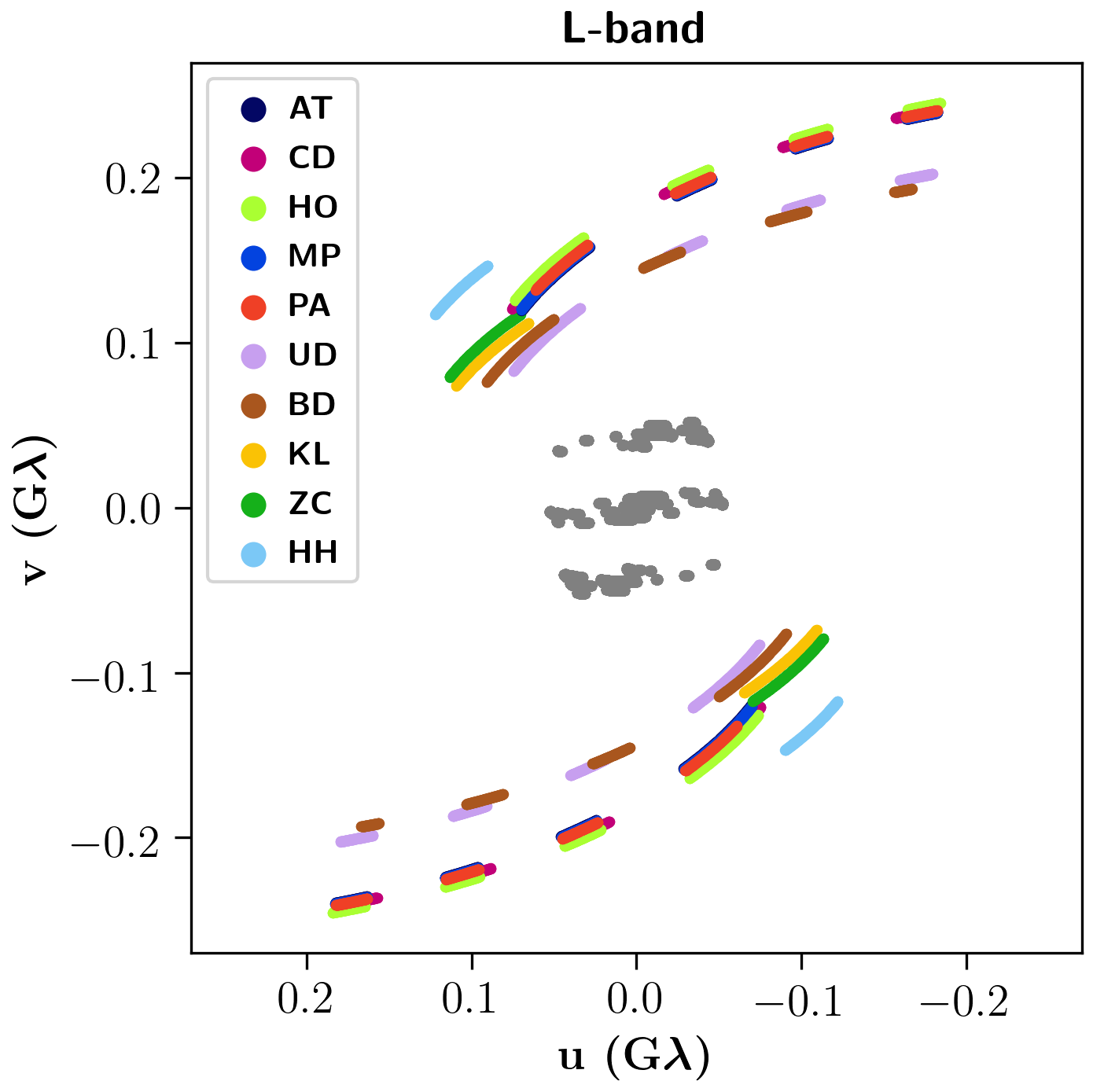}
\hspace{1em}
\includegraphics[width=0.48\textwidth]{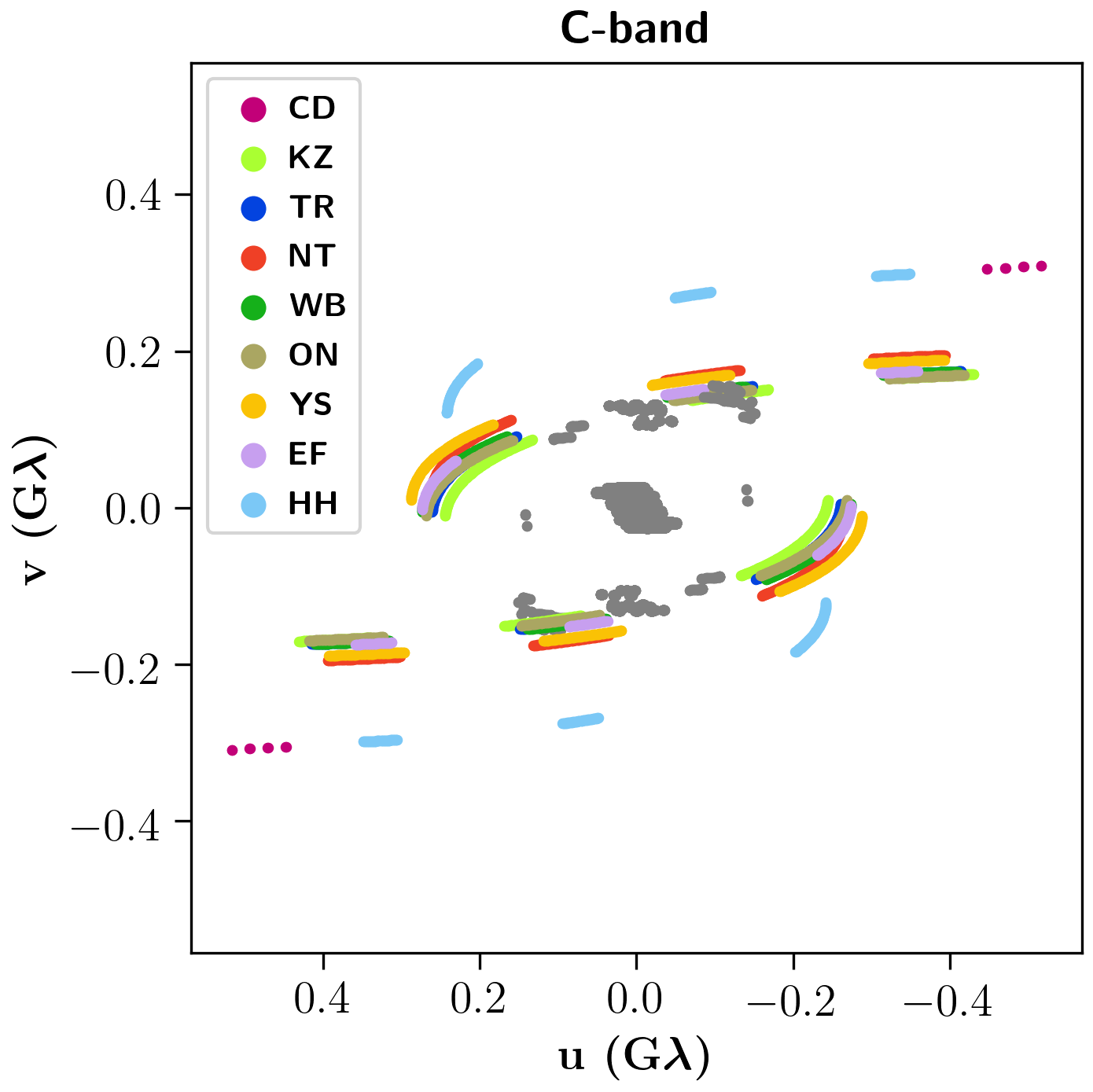}
\caption{The $u,v$-coverage for \emph{RadioAstron} observations at 1.6\,GHz (left panel) and 4.8\,GHz (right panel) presented in this work. The central bulge (in grey) of $u,v$-tracks span about one Earth diameter (ground stations baselines), while the ``wings'' (in colors) represent the \emph{RadioAstron} Space-baselines contribution. Only space-segments giving fringes are plotted, i.e., up to a maximum projected space--ground baseline of $\sim$4.5~ED at 1.6\,GHz and $\sim$3.3~ED at 4.8\,GHz.}
\label{uv-plots}
\end{figure*}

%%%%%%%%%%%%%%%%%%%%%%%%%%%%%%%%%%%%%%%%%%%%%%%%%%%%%%%%%%%%%%%%%%%%%%%%%%%%%%%%%%%%%%%%

\begin{figure*}
\centering
\includegraphics[width=\textwidth]{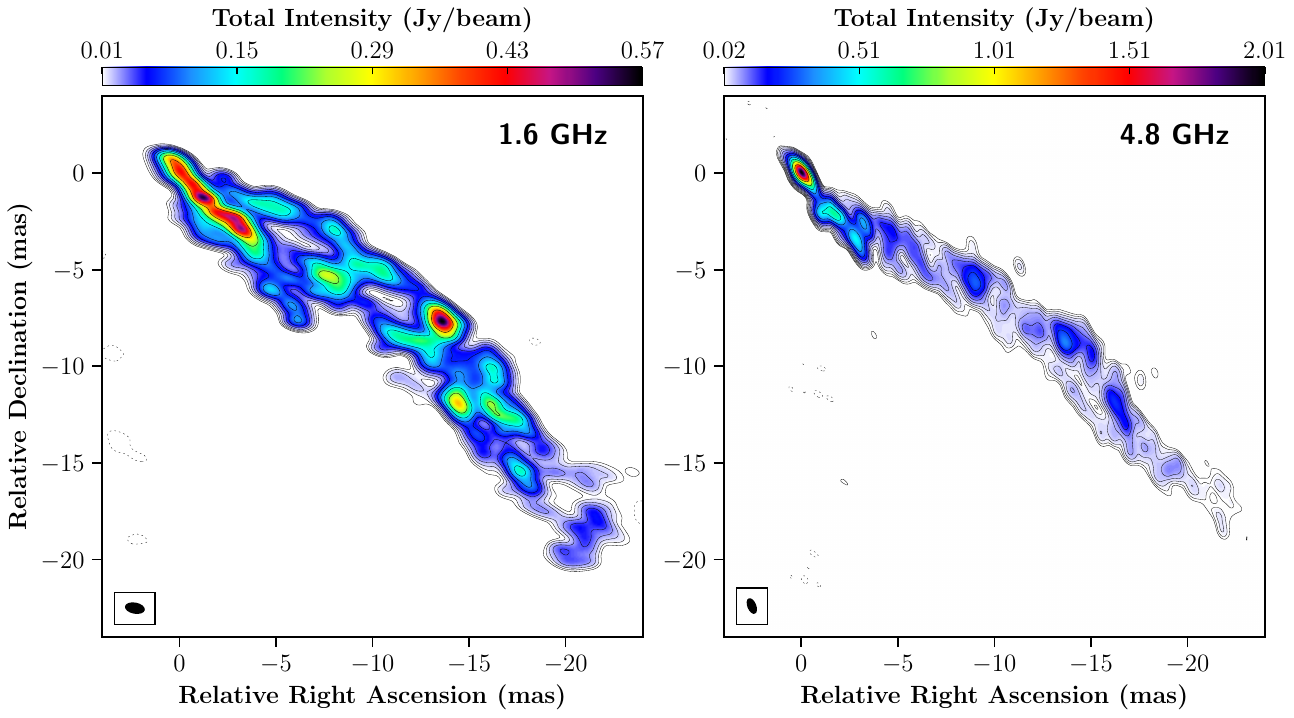}
\caption{\emph{RadioAstron} images of 3C\,273 at 1.6 GHz (left panel, June 2014) and 4.8 GHz (right panel, April 2014). The beam is shown on the lower-left corner: 1.04$\times$0.58 mas, P.A. 79.4$^\circ$ at 1.6 GHz, and 0.86$\times$0.46 mas, P.A. 22.0$^\circ$ at 4.8 GHz. The two lowest contour levels are $\pm$3 and $\pm$7 times the RMS of the image noise level (3$\times$3.5 mJy/beam at 1.6 GHz, 7$\times$2.2 mJy/beam at 4.8 GHz). Successive contours are drawn as $c_n = (3/2)\times c_{n-1}$ up to 90$\%$ of the total intensity peak (0.57 Jy/beam at 1.6 GHz, 2.01 Jy/beam at 4.8 GHz).}
\label{RA-images}
\end{figure*}

%%%%%%%%%%%%%%%%%%%%%%%%%%%%%%%%%%%%%%%%%%%%%%%%%%%%%%%%%%%%%%%%%%%%%%%%%%%%%%%%%%%%%%%%

\section{\emph{RadioAstron} observations and data processing}

Observations at 1.6\,GHz were performed on 2014 June 13, under project code {\tt{GA030F}} for ground antennas and {\tt{raks04f}} for RA. The array was composed of antennas in Russia (Kalyazin 64\,m, Badary 32\,m, Zelenchukskaya 32\,m), Japan (Usuda 64\,m), Australia (ATCA 5$\times$22\,m, Ceduna 30\,m, Hobart 26\,m, Mopra 22\,m, Parkes 64\,m), New Zealand (Warkworth 12\,m) and South Africa (Hartebeesthoek 26\,m). The tracking station for RA was Pushchino for the entire experiment. The total observing time was 9 hours (05--14 UT). RA participated with 10 scans, 14.5 minutes each, for a total of 2.4 hours on target, covering the space-baselines between 1.6 and 4.5 Earth Diameters (ED). Observations at 4.8\,GHz were performed on 2014 April 30 with the project code {\tt{GL038F}} for the ground array and {\tt{raks05d}} for RA. The antennas composing the ground array were in Russia (Kalyazin 64\,m), Australia (Ceduna, Hobart), South Africa (Hartebeesthoek), USA (Mauna Kea 25\,m), and Europe (Effelsberg 100\,m, Noto 32\,m, Onsala 25\,m, Torun 32\,m, Yebes 40\,m, Westerbork 11$\times$25\,m). The tracking station for RA was Pushchino for the entire experiment. The total observing time was 12 hours (10--22 UT). RA participated with 27 scans, 9.5 minutes each, for a total of 4.3 hours on target, covering the space-baselines between 0.9 and 3.3 ED. \autoref{uv-plots} presents $uv$-coverage in the two RA experiments discussed in this work. The data from both observing bands were processed at the MPIfR correlator, making use of the RA-dedicated version of {\tt DiFX} software VLBI correlator \citep{2016Galax...4...55B}. Fringe-fitting at the correlator stage was performed using the largest available antennas as references for each experiment (ATCA, Parkes, Effelsberg), first setting the clock value for the ground array antennas, and then searching for signal in each RA scan: this allowed us to have a first-order solution for each space-ground scan that could later be refined through baseline-stacking (see below for details). For scans giving no fringes, we applied extrapolated clock values from the successful part of the experiment. 

The following data reduction strategy in {\tt AIPS}\footnote{\href{http://www.aips.nrao.edu/index.shtml}{http://www.aips.nrao.edu/index.shtml}}, proven to be successful for RA AGN imaging projects, was adopted for both experiments. First, the {\em a priori} amplitude calibration was applied using the values for the antenna gains and system temperature measured at each antenna during the observations. A parallactic angle correction was applied to the ground array antennas to account for the axis rotation of the antenna feeds with respect to the target source. The data were then fringe-fitted.  The ground array data were fringe-fitted first, then the SVLBI baselines, using stacked solutions of the ground baselines and a model of the source (baselines stacking within the {\tt FRING} task in {\tt AIPS}). This procedure was repeated scan by scan for the SVLBI baselines in order to use a more accurate model describing the source structure. The solution interval was set to 2~min for ground array observations and to 4~min for the SVLBI scans, adopting a SNR threshold of 5. At 1.6\,GHz, SVLBI fringes were found on baselines up to $\sim$4.5 Earth Diameters (ED), while at 4.8\,GHz up to $\sim$3.3~ED. 

Finally, we imaged the calibrated data in {\tt Difmap}\footnote{\href{ftp://ftp.astro.caltech.edu/pub/difmap/difmap.html}{ftp://ftp.astro.caltech.edu/pub/difmap/difmap.html}} \citep{1997ASPC..125...77S}. First, we flagged RA scans for which SVLBI fringes gave a SNR$<$5 in {\tt AIPS}. Visibilities were then averaged over 30 seconds at 1.6 GHz, while over 10 seconds at 4.8 GHz, and the standard uniform weighting scheme ($\mathrm{uvw}=2,-1$) was applied at both frequencies. The source model was built through the standard iterative {\tt CLEAN}ing and phase self-calibration technique, adopting solution intervals equal to the data averaging time. The feasibility of phase self-calibration was assured by the high SNR of the visibilities, and the abundance of baselines. Once a satisfactory model was assessed, amplitude self-calibration was performed as well. The shortest amplitude self-calibration solution interval was chosen to be 1 minute for both bands. The final images angular resolution and RMS were: 1.04$\times$0.58~mas and 3.5~mJy/beam at 1.6~GHz, while 0.86$\times$0.46~mas and 2.2~mJy/beam at 4.8~GHz.

%%%%%%%%%%%%%%%%%%%%%%%%%%%%%%%%%%%%%%%%%%%%%%%%%%%%%%%%%%%%%%%%%%%%%%%%%%%%%%%%%%%%%%%%

\section{Results}

\subsection{Limb-brightened jet emission}

The RA observations presented in this work allowed us to reach an unprecedented angular resolution at 1.6 GHz for this source, and to detect for the first time a limb-brightened jet in 3C\,273 (\autoref{RA-images}, left panel). The beam in this image obtained with a uniform weighting scheme is 1.04$\times$0.58 mas, with a position angle (PA) of 79.4$^\circ$ and image RMS noise of 3.5 mJy/beam. As clearly visible, the structure of the jet is dominated by an elongated structural feature unresolved in the direction orthogonal to the jet for the inner 5~mas from the core, while two distinct emission trails are visible along the rest of the jet, extending towards the south-west for further $\sim$20~mas. A brighter region is present at about 15~mas from the core, where the jet bends towards the south, that may be associated with a shock produced by the interaction of the jet with the ambient medium (cf. the cases of 4C\,41.17, \citealt{1997A&A...318...11G}, and J0906$+$6930, \citealt{2020An_NatComm}). This limb-brightened emission was not seen in previous observations of 3C\,273, most probably because of the lower angular resolution of previous images (see, e.g., the VLBA image, obtained at 1.7 GHz in 2011, reported in \citealt{2016ApJ...820L...9K}). 

\begin{figure*}
\centering
\includegraphics[width=\textwidth]{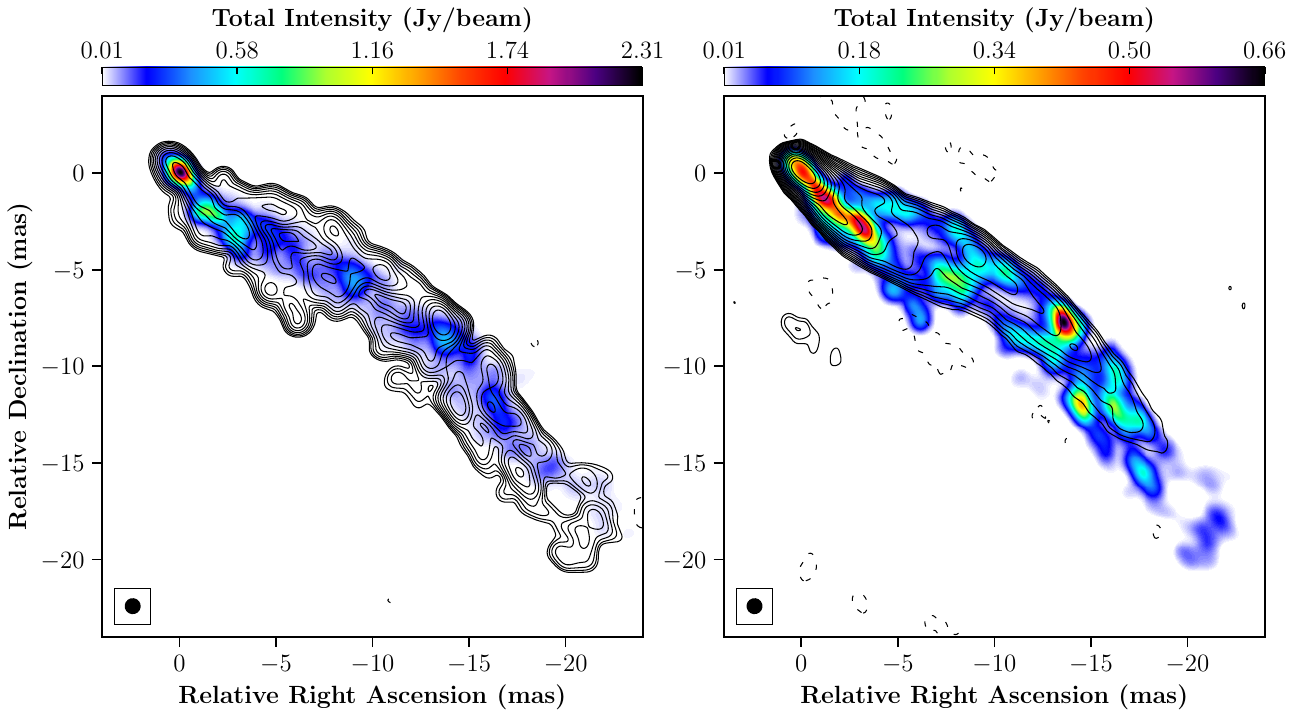} 
\caption{Left panel: the \emph{RadioAstron} images of 3C\,273 at 1.6\,GHz (contours) and 4.8\,GHz (in colors) obtained in 2014. Both are convolved with the MOJAVE circular beam (0.83$\times$0.83 mas). Right panel: the \emph{RadioAstron} 1.6\,GHz image (in colors) with the stacked image from MOJAVE (contours), both convolved with the MOJAVE circular beam of the same size as in the left panel. The two lowest contour levels are $\pm5$ and $\pm9$ times the RMS noise level: 5$\times$2.8\,mJy/beam at 1.6\,GHz and 9$\times$0.5\,mJy/beam at 15~GHz, respectively. Successive contours are drawn as $c_n = (3/2)\times c_{n-1}$ up to 90$\%$ of the total intensity peak (0.7 Jy/beam at 1.6 GHz, 6.1 Jy/beam at 15 GHz).}
\label{L+C}
\end{figure*}

%%%%%%%%%%%%%%%%%%%%%%%%%%%%%%%%%%%%%%%%%%%%%%%%%%%%%%%%%%%%%%%%%%%%%%%%%%%%%%%%%%%%%%%%

\begin{figure*}
\centering
\includegraphics[width=\textwidth]{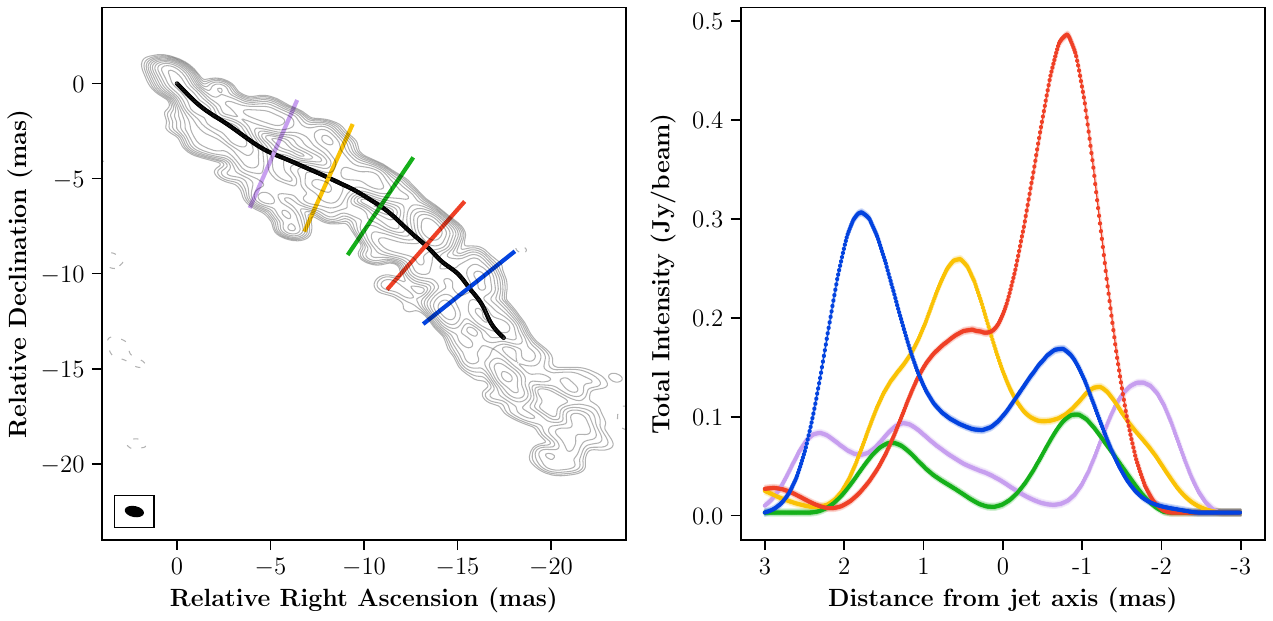} 
\caption{Example of transverse jet brightness profile (right) for different positions along the stream axis (left) at 1.6~GHz.}
\label{profile}
\end{figure*}

%%%%%%%%%%%%%%%%%%%%%%%%%%%%%%%%%%%%%%%%%%%%%%%%%%%%%%%%%%%%%%%%%%%%%%%%%%%%%%%%%%%%%%%%

We further investigated the limb-brightening properties through a detailed tomography along the jet. The MOJAVE team produced stacked images of 373 jets, including 3C\,273, at 15\,GHz with a time span of 20 years \citep{2017MNRAS.468.4992P}. As demonstrated in that work, it is necessary to stack the images in order to properly map the full jet width at 15\,GHz. In \autoref{L+C} (right panel), we present the contours from the MOJAVE stacked image overlaid on our L-band image, with the purpose of comparing the different structure. A shift of 43$\times$55 pixels in RA and Dec, respectively, was applied to align the two images, corresponding to 2.15 mas eastward and 2.75 mas northward (see \autoref{Sec:L+C} for a full description of the method).
No evident limb-brightened emission is detected in the MOJAVE data. Furthermore, the emission is not symmetric along the full length of the observed jet: the first $\sim$5\,mas from the core shows an enhanced emission in the southern region, while this shifts to the northern side from $\sim$5 to $\sim$10\,mas. Remarkably, the cores and the innermost $\sim$5\,mas jet emission are consistent in the two images, and the limb-brightening lies at the edges of the MOJAVE contours. Overall, the MOJAVE intensity distribution, including the jet curvature, is in agreement with our SVLBI results. 

We used the jet ridge-line from MOJAVE observations to measure the jet profile along slices perpendicular to it in our RA image, for a total of 440 profiles of 6\,mas width. The first 72 profiles, corresponding to the inner 3\,mas, do not show a clear double-peaked structure. Down the flow at larger distances, a double structure is visible, but the two peaks are very close to each other. This structure is observed between the inner $\sim3$\,mas and $\sim4$\,mas. Beyond 4\,mas, the double-peaked structure is more prominent, showing a clear limb-brightening. In \autoref{profile}, we present five representative jet profiles, clearly showing the double peak resulting from the limbs. The latter can be roughly four times brighter than the central region of the jet. Moreover, as seen in the MOJAVE stacked image, the southern region is brighter in the first few mas from the core, while the northern side is stronger in the second half of the jet. 

%%%%%%%%%%%%%%%%%%%%%%%%%%%%%%%%%%%%%%%%%%%%%%%%%%%%%%%%%%%%%%%%%%%%%%%%%%%%%%%%%%%%%%%%

\begin{figure*}
\centering
\includegraphics[width=\textwidth]{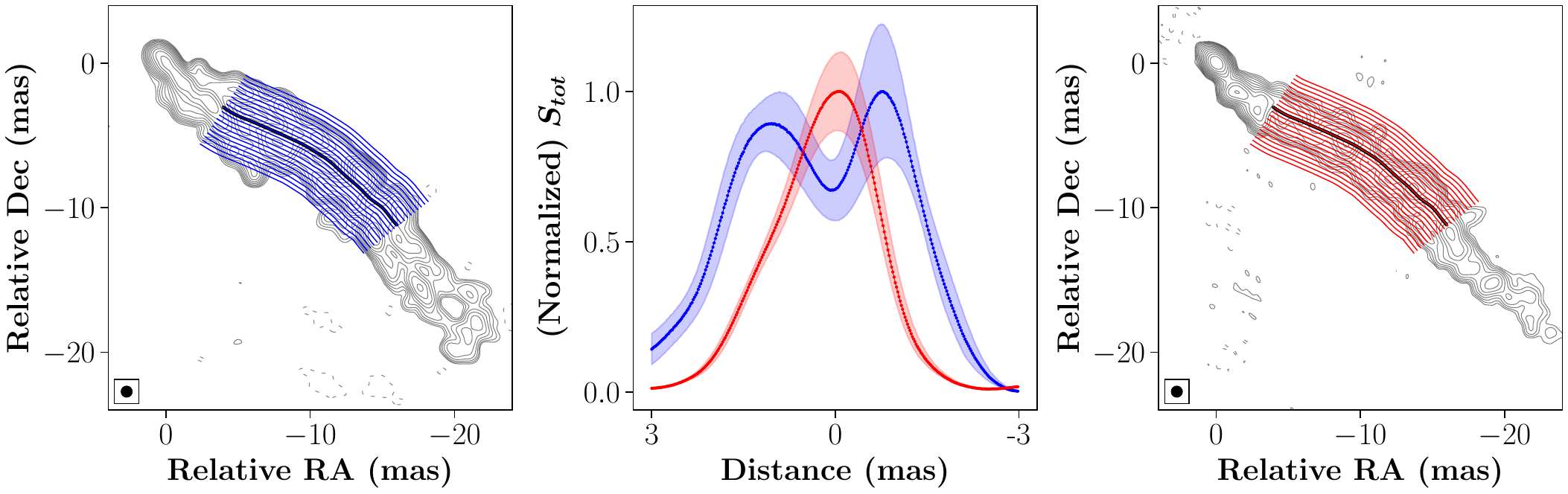}
\caption{The 300 jet streamlines used to calculate the integrated flux density along the flow, parallel to the jet axis from MOJAVE (central line in black), overplotted on the 1.6\,GHz (left panel) and 4.8\,GHz (right panel) \emph{RadioAstron} contours maps (both convolved with the MOJAVE beam). The starting point is set at 4~mas from the core, where the spine/sheath structure becomes evident. Central panel: integrated flux density, versus distance from jet axis, calculated along jet streamlines for the 1.6\,GHz (left) and 4.8\,GHz (right) \emph{RadioAstron} images. The profile uncertainty is reported as a shaded area.}
\label{fluid}
\end{figure*}

%%%%%%%%%%%%%%%%%%%%%%%%%%%%%%%%%%%%%%%%%%%%%%%%%%%%%%%%%%%%%%%%%%%%%%%%%%%%%%%%%%%%%%%%

\subsection{Evidence of a spine/sheath structure along the jet}\label{Sec:L+C}

Given the results from the 1.6\,GHz observations described above, we considered the RA 4.8\,GHz image of the same source, obtained less than 2 months earlier, in order to compare the two jet structures. The 4.8\,GHz image is presented in \autoref{RA-images}: an angular resolution of 0.86$\times$0.46 mas was obtained (uniform weighting) with a beam PA of 22$^\circ$. The average image RMS noise is $\sim$2.2\,mJy/beam. In this image, a single stream is visible for the whole jet, with no indication of any limb-brightening, contrary to what is seen at 1.6\,GHz. Remarkably, the jet curvature is the same as reconstructed in the 1.6\,GHz image, and the  brighter regions are also in agreement between the two maps. 

To better compare the different jet morphologies visible in the two images, we superimposed the 1.6\,GHz map with the one at 4.8\,GHz, restored with a matched circular beam equal to the MOJAVE one (0.83$\times$0.83 mas). Image registration was performed via a cross-correlation analysis of the total intensity maps (see \citealt[][and references therein]{2016ApJ...817...96G}). In particular, we considered the inner jet spine only, which looks similar in the two images, covering the first $\sim$5 mas. Several spectral index maps were produced, adopting a different shift between the two maps, to identify by visual inspection the one showing the smallest gradients across the jet width (see \citealt{2019MNRAS.485.1822P} for more details about this method). The final shift adopted was 36$\times$48 pixels in RA and Dec, respectively, for the 4.8\,GHz image, corresponding to 1.8 mas eastwards and 2.4 mas northward.
The resulting image is presented in \autoref{L+C} (left panel), where 1.6 \,GHz is represented in contours and 4.8\,GHz in colors. It is evident that the emission at 4.8\,GHz falls between the two limbs detected at 1.6\,GHz, suggesting that the former traces the jet spine, while the latter the sheath. Although dominance of the jet spine or the jet sheath has been observed in other sources, for the first time here we detect the two structures at frequencies so closely spaced (1.6~GHz vs. 4.8~GHz).

Finally, in order to quantify the prominence of the limbs and the spine at the two frequencies, we traced 300 lines parallel to the jet axis, and calculated the integrated flux density along the lines for both frequencies. To compute the profiles we cut the jet orthogonally to the MOJAVE ridgeline. Since the ridgeline is curved, a line is first fitted between two adjacent points along the ridgeline, then a cut perpendicular to this line is made. We repeat this process for the whole length of the ridgeline and derive the flux density along each cut. Finally, to estimate the profile of the jet flux density between the two sides at 1.6\,GHz and 4.8\,GHz, we have traced parallel lines to the jet axis (i.e., the MOJAVE ridgeline), also called fluid lines. We created a total of 300 parallel lines, covering the entire jet width, and summed the flux density along the lines. The fluid lines used to integrate the flux density along the flow, with estimated integrated profiles, are shown in \autoref{fluid}. 
Uncertainties have been computed for each streamline as: 
\begin{equation}
\sigma_p = \frac{N}{\sqrt{N_\text{beams}}} \sigma_S,
\end{equation}
where $N$ and $N_\text{beams}$ are the number of pixels and beams along the streamline, respectively, while $\sigma_S$ is the standard deviation of the flux density values along it. This formulation, obtained through the covariance matrix calculation, allow us to take into account the correlation of pixels' flux densities within the beam.
We found again a tangential double-peaked profile at 1.6\,GHz, and a single-peaked one at 4.8\,GHz (see \autoref{fluid}), confirming the spine/sheath jet structure. 

It should be noted that the double helical structure reported in the C-band image of 3C\,273 obtained from the VSOP observation \citep{2000AdSpR..26..669L,2001Sci...294..128L} may remain undetected in our C-band image, owing to an $\approx 3.5$ times smaller beam and a factor of $\approx 2.5$ lower dynamic range of the RA C-band image presented here. This results in a sensitivity to weak and extended emission almost an order of magnitude lower, which may preclude effective detection of the transverse structure of the flow in the RA image at distances along the jet larger than $\approx 5$~mas.

%%%%%%%%%%%%%%%%%%%%%%%%%%%%%%%%%%%%%%%%%%%%%%%%%%%%%%%%%%%%%%%%%%%%%%%%%%%%%%%%%%%%%%%%

\begin{figure*}
\centering
\includegraphics[width=\textwidth]{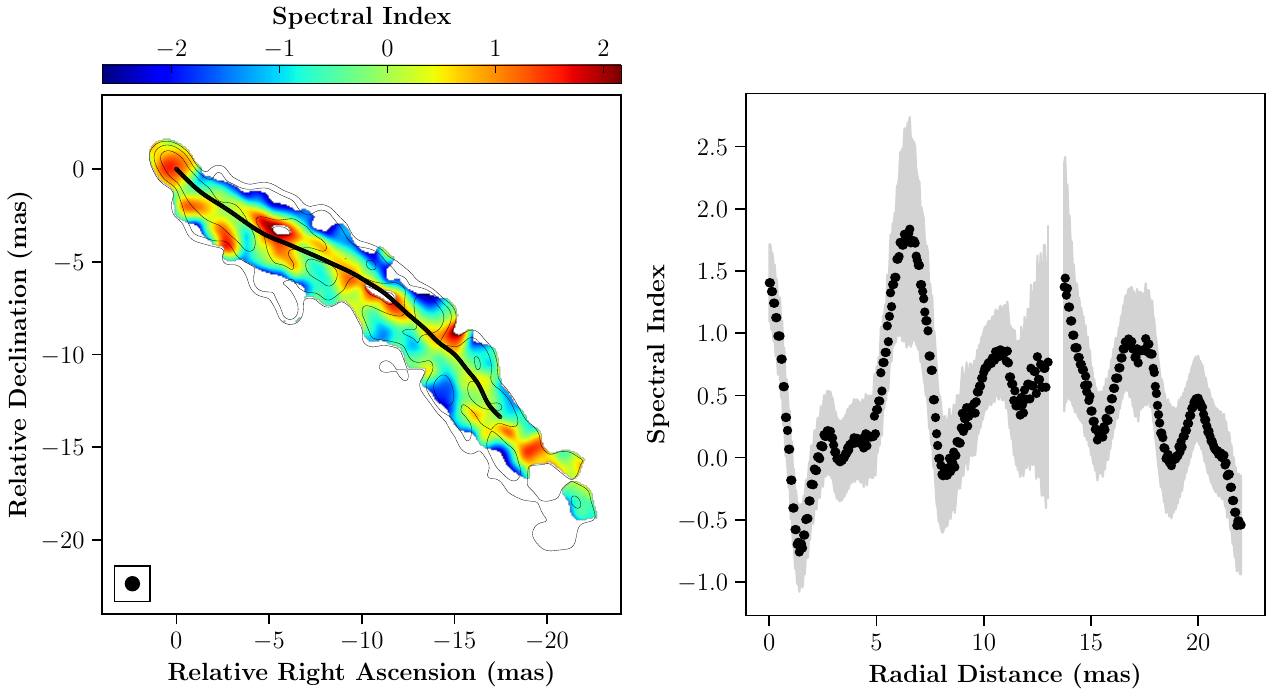}
\caption{Left panel: spectral index map obtained from the \emph{RadioAstron} images of 3C\,273 at 1.6\,GHz and 4.8\,GHz, both restored considering only the common UV-range 2.5-340 M$\lambda$, and convolved with the MOJAVE circular beam (0.83$\times$0.83 mas). The map is plotted over the 1.6 GHz image shown in contours. Pixel values below 5$\sigma$ for both the 1.6 GHz and 4.8 GHz images were blanked. Right panel: spectral index value and its uncertainty along the jet axis, plotted as a function of the radial distance from the core. The adopted convention for the spectral index $\alpha$ is $S\propto\nu^\alpha$.}
\label{SI}
\end{figure*}

\section{Possible physical factors concurring in the observed jet structure}\label{Sec:factors}

%%%%%%%%%%%%%%%%%%%%%%%%%%%%%%%%%%%%%%%%%%%%%%%%%%%%%%%%%%%%%%%%%%%%%%%%%%%%%%%%%%%%%%%%
%%%%%%%%%%%%%%%%%%%%%%%%%%%%%%%%%%%%%%%%%%%%%%%%%%%%%%%%%%%%%%%%%%%%%%%%%%%%%%%%%%%%%%%%

\subsection*{Propagating structures along the jet}

\cite{kups9379} reported an oscillation in the jet
direction of 3C\,273 between the previous VSOP observation \citep{2001Sci...294..128L}
and the RA observation, separated by seventeen years. The
author reported an oscillation velocity of $\simeq0.5\,c$, and a pattern
speed of  $(0.070\pm0.016)\,c$. These velocities, clearly smaller than
the flow speed revealed by, e.g., the strong brightness asymmetry, show
that the observed structures are caused by waves that propagate through
the jet (i.e.\ helical patters, see \citealt{2012ApJ...749...55P,2015ApJ...803....3C,2019A&A...627A..79V}). The wavelengths that can be derived from the RA observations reported there coincide with those given in \cite{2001Sci...294..128L}, plus a further, longer wavelength of $\simeq 50\,{\rm mas}$, due to the sensitivity achieved at larger scales. 

In this respect, the limb brightening observed in this work can be caused by a combination of the off-axis pressure enhancements caused by the helical and elliptical waves (see \citealt{2001Sci...294..128L}) and the interaction of the jet with its environment as it
oscillates, in the same way as reported by \cite{2018ApJ...855..128W} for the
case of M87. The compression of the gas and the magnetic lines due to
the medium resistance to jet expansion can cause the observed rise in
emissivity. 

%%%%%%%%%%%%%%%%%%%%%%%%%%%%%%%%%%%%%%%%%%%%%%%%%%%%%%%%%%%%%%%%%%%%%%%%%%%%

\subsection*{Velocity stratification in the jet flow}

The observed limb-brightening of the RA image at 1.6\,GHz requires a stratification across the jet width in the flow density, internal energy, magnetic field, and/or bulk flow velocity. 
This conclusion is further supported by the 1.6/4.8\,GHz spectral index distribution plotted in \autoref{SI}. This distribution also shows a clear transverse structure in which the spectral index changes between $-$0.2 and $+$1.7 along the jet ridge, while it gradually drops below $-$0.7 towards the jet edges.
A stratified jet, with a steep increase in the particle and/or magnetic field energy towards the jet edges can naturally explain the limb brightening and the spectral index distribution, as shown for instance in \cite{2019ApJ...877...19O}. %[Alternatively we could try to cite some jet formation simulations that show a denser jet sheath. RMHD simulations also show some stratification in B, but this doesn't necessarily yield a limb-brightening.] 
Transverse stratification of the jet in 3C\,273 has indeed been suggested from numerical simulations modelling the Kelvin-Helmholtz instability developing in the flow \citep{2006A&A...456..493P}.
Numerical simulations also predict a faster jet spine (or beam), surrounded by a slower and denser sheath (see, e.g., the previously mentioned two-fluid models, developed by various authors starting in the 1980s). For a given viewing angle, $\theta$, we can compute the Lorentz factor that maximizes the Doppler boosting, which is given by $\Gamma_{\max}=sin^{-1}\theta$. Jets with a bulk Lorentz factor at the jet spine significantly larger than $\Gamma_{\max}$ will show a limb-brightening, while those with a value similar or smaller than this will show a spine brightening instead.

For 3C\,273, several kinematic measurements \citep{2005AJ....130.1418J,2006A&A...446...71S,2019ApJ...874...43L} show apparent speeds, $\beta_\mathrm{app}$  in the range of 4--16\,$c$, with a median value at around 9\,$c$. These measurements support estimates of the viewing angle $\theta\approx 7^{\circ}$--$19^{\circ}$ and Lorentz factor $\Gamma \approx 6$--$16$. The respective median values of $\theta = 14^{\circ}$ and $\Gamma_\mathrm{b} \ge 10$ can be taken as plausible estimates for the spine. The properties of the slower flow can be represented by the Lorentz factor $\Gamma_\mathrm{s}=2.1\pm 0.4$ estimated from the analysis of internal structure of the jet emission at 1.6 GHz \citep{2001Sci...294..128L}, noting also that the viewing angle estimate of $15^{\circ}\pm 3^{\circ}$ obtained in that analysis agrees well with the median value of the kinematic estimates made for the spine. For these parameters of the spine and flow, the differential Doppler boosting alone should provide limb brightness enhancements by a factor of $\ge 1.5$.
Hence, the velocity stratification in the jet flow, with a progressive deceleration towards the jet edges, may indeed produce the limb-brightening emission seen in the 1.6\,GHz RA image. 

%%%%%%%%%%%%%%%%%%%%%%%%%%%%%%%%%%%%%%%%%%%%%%%%%%%%%%%%%%%%%%%%%%%%%%%%%%%%
\subsection*{Plasma stratification across the jet}

Taken alone, the velocity stratification discussed above would equally produce a limb-brightened structure also at 4.8\,GHz, which is obviously inconsistent with the observed spine brightening observed at this frequency. For the jet to become spine-brightened at 4.8\,GHz, the magnetic field, the density of the emitting plasma, or the electron energy distribution must be varying across the jet. For instance, if higher energy electrons are concentrated around the jet spine, it would cause the emission to become more spine-brightened at progressively higher frequencies. This should lead to a clear stratification in the spectral index across the jet width, as visible in \autoref{SI}. A similar effect can be achieved by variable opacity across the jet induced by a stratified magnetic filed and/or particle density. To provide basic quantitative estimates, we can assume that the flow has a sharp transition from the spine (beam) with the radius $R_\mathrm{b}$ to the sheath with the outer radius $R_\mathrm{s}$.

Assuming a power-law electron energy distribution $N(E) = N_0 E^{-s}$ for both these regions, we can write a generic proportionality  \citep{1985ApJ...298..114M,1999ApJ...521..509L} for the synchrotron flux density received at a frequency $\nu$ from either of these two regions,
$S_\nu \propto N_0\, B^{1-\alpha}\, V_\mathrm{em}\, \delta^{2-\alpha}\, \nu^{\alpha}$, 
where $N_0$ is the particle density, $B$ is the magnetic field strength, $V_\mathrm{em}$ is the emitting volume, $\delta$ is the Doppler factor, and $\alpha$ is the spectral index of the emission. Denoting these quantities with subscripts $b$ for the spine (beam) and $s$ for the sheath, we can obtain the ratio $S_\mathrm{\nu,b}/S_\mathrm{\nu,s}$ of the flux densities of the spine and the sheath. From the profiles in the middle panel of \autoref{fluid}, we estimate $S_\mathrm{\nu,b}/S_\mathrm{\nu,s}\approx 0.7$ and $R_\mathrm{s} \approx 1.9\,R_\mathrm{b}$. For an axisymmetric jet structure, the latter translates into the emitting volume ratio $V_\mathrm{em,b}/V_\mathrm{em,s} = R_\mathrm{b}^3/(R_\mathrm{s}^3-R_\mathrm{b}^2 R_\mathrm{s}) \approx 0.2$. The kinematic measurements discussed above yield $\delta_\mathrm{b}\approx 2.8$ and $\delta_\mathrm{s}\approx 3.2$ for the Doppler factors in the spine and the sheath, respectively.

Two scenarios can be considered for the spectral indices of the spine and the sheath. In the first scenario, the spine and the sheath contain plasma with the same electron energy distribution. For this scenario, we set both spectral indices to $-$0.7, which corresponds to a commonly assumed $s=2.4$ for the electron energy distribution. In the second scenario, the two regions contain different plasmas, and the respective spectral indices also differ. The spectral index image and the ridge line profile of the spectral index in \autoref{SI} can be used to estimate $\alpha_\mathrm{b}\approx -0.2$ and $\alpha_\mathrm{s}\approx -0.8$. The $\alpha_\mathrm{b}$ is estimated from the spectral index of the underlying emission in the spine, while the stretches of higher spectral index observed in that profile must correspond to relativistic shocks propagating through the spine. Indeed, mapping the synchrotron turnover frequency has shown that it is larger than 5\,GHz in the shocked regions \citep{1997VA.....41..253L}, and therefore these regions should have a positive 1.6/4.8\,GHz spectral index.

With all the estimates discussed above, we obtain for the first scenario
$S_\mathrm{\nu,b}/S_\mathrm{\nu,s} \approx 0.07\, (N_\mathrm{0,b}/N_\mathrm{0,s})\, (B_\mathrm{b}/B_\mathrm{s})^{1.7}\,.$
In order to reconcile this relation with the measured flux density ratio of 0.7, the condition $(N_\mathrm{0,b}/N_\mathrm{0,s})\, (B_\mathrm{b}/B_\mathrm{s})^{1.7} = 9.8$ needs to be satisfied. In stratified jets \citep{2020ARA&A..58..407D,2021NewAR..9201610K}, the particle density is expected to be lower in the spine, while rising approximately linearly with increasing distance, $r$, from the jet axis \citep{2016ApJ...831..163M} and reaching a maximum inside the sheath or in the transition region. As the onset of the observed transverse stratification occurs at distances $\gtrsim$\,30 pc from the observed jet origin, the magnetic field there is expected to generate a relatively strong poloidal component (through shear, see \citealt{2018A&A...610A..32B,2020A&A...644A..85S}) dropping off roughly $\propto r^{-2}$. In this case, assuming $N_\mathrm{0,b} = (R_\mathrm{s}/R_\mathrm{b})\,N_\mathrm{0,s} = 0.5$ (for linearly increasing particle density), we get $B_\mathrm{b}/B_\mathrm{s} \approx 5.6$. This implies $B\propto r^{-2.7}$, which is slightly steeper than the theoretical expectation but can be still viewed as a reasonable agreement. Requiring $B\propto r^{-2}$ results in $N \propto r^{0.1}$, which would imply that both the spine and the sheath are located near the layers of the flow with the maximum particle density. 

For the second scenario, with different electron energy distributions in the spine and the sheath, we obtain the ratio
$S_\mathrm{\nu,b}/S_\mathrm{\nu,s} \approx 0.09\, (N_\mathrm{0,b}/N_\mathrm{0,s})\, (B_\mathrm{b}^{1.2}/B_\mathrm{s}^{1.8})\,.$
If one requires that the measured ratio of 0.7 should be explained solely by changes in the electron energy distribution (hence without any stratification in $N$ and $B$), the latter ratio implies $B_\mathrm{b}=B_\mathrm{s}=30$\,mG. At 1\,pc distance from the jet origin, the magnetic field strength $B_\mathrm{1pc}=0.3$--1.0\,G has been estimated \citep{2017MNRAS.468.4478L}. With this estimate and for a poloidally dominated field, one would expect field strengths of $\approx4$--10\,mG, averaging to $B_\mathrm{avg}\approx 7$\,mG, over the region contributing to the transverse profiles plotted in \autoref{fluid}. This is somewhat lower than the field strength of 30\,mG needed to explain the measured spine-sheath flux density ratio without stratification of the magnetic field and particle density.

Allowing for the stratification (again, assuming $N(r)\propto r$ between the spine and the sheath), we get for the magnetic field ratio $B_\mathrm{b}^{1.2}/B_\mathrm{s}^{1.8} = 4.3$. In this case, satisfying the expectation $B\propto r^{-2}$ for the poloidal field, we get $B_\mathrm{b}=4.1$\,G and $B_\mathrm{s}=1.1$\,G, which are unrealistic for the flow at these distances. Equating $B_\mathrm{b}$ to $B_\mathrm{avg}=7$\,mG yields $B_\mathrm{s}=16$\,mG which is difficult to reconcile with the poloidal field. For a toroidally dominated field, estimating the $B_\mathrm{b}$ from $B_\mathrm{1pc}$ gives a field strength of $\approx 70\,\mu$G and the respective $B_\mathrm{s}\approx 750\,\mu$G. These values both are too low and imply exceedingly steep field gradients across the jet. Altogether, introducing two different electron distributions for the spine and the sheath makes it more difficult to achieve an agreement with the measured flux density ratio. We can therefore conclude that moderately stratified particle density and magnetic field described in the first scenario can be sufficient for explaining the observed limb-brightened jet morphology at 1.6\,GHz and spine-brightened structure at 4.8\,GHz.

\subsection*{Potential instrumental effects}

Significant gaps in the \emph{u,v}-plane can potentially generate artifacts in the final image. The experimental setup of SVLBI observations can easily incur in such issue, when performed with a space antenna at a distance larger than $\sim$1 ED from ground. However, depending on the brightness, extension, and morphology of the source under study, the potential artifacts have often a limited impact on the final images. Through a multi-epoch analysis of the jet morphology in S5\,0836+710, \cite{2012ApJ...749...55P} and \cite{2019A&A...627A..79V} discussed how \emph{u,v}-coverage should not introduce relevant differences on the observed jet structure, but only minor ripples along the ridge lines. Indeed, they found the latter to be consistent among different epochs of observations, and instrumental setups. The features discussed in the present work are prominent ($\mathrm{SNR}>100$), with an extension of several beams, and visible as long waves along the jet ($\sim$10 mas), so we can reasonably exclude an instrumental effect due to gaps in \emph{u,v}-coverage. In addition, it has been shown that spectral indices (and then flux distribution) obtained from observations with uneven \emph{u,v}-coverage can be trusted (i.e. have an accuracy >90\%) for a sufficient pixel SNR (>5), and when within 10--15 mas from the phase-center \citep{1998A&AS..132..261L}. Both those conditions are satisfied by the structures presented in this work.

%%%%%%%%%%%%%%%%%%%%%%%%%%%%%%%%%%%%%%%%%%%%%%%%%%%%%%%%%%%%%%%%%%%%%%%%%%%%%%%%%%%%%%%%
%%%%%%%%%%%%%%%%%%%%%%%%%%%%%%%%%%%%%%%%%%%%%%%%%%%%%%%%%%%%%%%%%%%%%%%%%%%%%%%%%%%%%%%%

\section{Conclusions}
We have presented an analysis of RA images at 1.6\,GHz and 4.8\,GHz for 3C\,273, the former being the highest angular resolution image to date for this source at this frequency. Our findings can be summarized as follows: 

\begin{itemize}
\item For the first time, a limb-brightened emission is evident in the image at 1.6\,GHz, showing an enhanced emission for the two edges of the jet, starting from about 4\,mas from the core and following the whole jet extension. 

\item Conversely, at 4.8\,GHz only a single stream is detected, consistently located between the edges of the 1.6\,GHz jet. This is confirmed by the jet profile drawn from the integrated flux density along the jet streamlines, indicating a double-peaked profile at 1.6\,GHz and a single-peaked one at 4.8\,GHz. 

\item The observed morphology is indicative of a spine/sheath structure in the jet. This can be explained in terms of the following, concurring, physical factors: 1) helical patterns propagating along the jet, similar to the ones reported in previous Space-VLBI observations of this source \citep{2001Sci...294..128L}; 2) a velocity stratification in the jet flow, with a faster jet spine and a slower and denser sheath; 3) plasma stratification across the jet, necessary to produce the noticeable morphological differences between the two observed close frequencies. This stratification scenario is supported by the spectral index gradient measured across the jet, and consistent along its extension.
\end{itemize}

A more detailed quantitative consideration through dedicated general relativistic magneto-hydrodynamical numerical simulations, able to reproduce the spine/sheath structure seen in these RA observations of 3C\,273, will be published in a future work.

%%%%%%%%%%%%%%%%%%%%%%%%%%%%%%%%%%%%%%%%%%%%%%%%%%%%%%%%%%%%%%%%%%%%%%%%%%%

\begin{acknowledgements}
We thank the anonymous referee for useful comments.
JLG and AF acknowledge financial support from the Spanish Ministerio de Econom\'{\i}a y Competitividad (grants AYA2016-80889-P, PID2019-108995GB-C21), the Consejer\'{\i}a de Econom\'{\i}a, Conocimiento, Empresas y Universidad of the Junta de Andaluc\'{\i}a (grant P18-FR-1769), the Consejo Superior de Investigaciones Cient\'{\i}ficas (grant 2019AEP112), and the State Agency for Research of the Spanish MCIU through the Center of Excellence Severo Ochoa award for the Instituto de Astrof\'{\i}sica de Andaluc\'{\i}a (SEV-2017-0709). APL, YYK, and ABP were supported by the Russian Science Foundation (project 20-62-46021). TS was partly supported by the Academy of Finland projects 274477 and 315721. MP acknowledges the support by the Spanish  Ministerio de Ciencia e Innovaci\'{o}n (MICINN) under grant PID2019-105510GB-C31. MP and JMM acknowledge financial support from the Spanish Ministry of Science through Grants PID2019-107427GB-C33 and AYA2016-77237-C3-3-P, and from the Generalitat Valenciana through grant PROMETEU/2019/071. JMA was supported by by the German Research Foundation grant
HE5937/2-2. LIG acknowledges support by the CSIRO Distinguished Visitor Programme. The \emph{RadioAstron} project is led by the Astro Space Center of the Lebedev Physical Institute of the Russian Academy of Sciences and the Lavochkin Scientific and Production Association under a contract with the Roscosmos State Corporation, in collaboration with partner organizations in Russia and other countries.
This publication has received funding from the European Union's Horizon 2020 research and innovation programme under grant agreement No 730562 [RadioNet]. 
This paper includes data observed with the 100-m Effelsberg radio-telescope, which is operated by the Max-Planck-Institut für Radioastronomie in Bonn (Germany). 
The National Radio Astronomy Observatory is a facility of the National Science Foundation operated under cooperative agreement by Associated Universities, Inc.
The European VLBI Network is a joint facility of independent European, African, Asian, and North American radio astronomy institutes. 
The Long Baseline Array is part of the Australia Telescope National Facility which is funded by the Australian Government for operation as a National Facility managed by CSIRO.
This research made use of Python ({\tt \href{http://www.python.org}{http://www.python.org}}), Numpy \citep{5725236}, Pandas \citep{mckinneyprocscipy2010}, and Matplotlib \citep{Hunter:2007}. We also made use of Astropy ({\tt \href{http://www.astropy.org/}{http://www.astropy.org}}), a community-developed core Python package for Astronomy \citep{Astropy:2013, Astropy:2018}.
\end{acknowledgements}

%%%%%%%%%%%%%%%%%%%%%%%%%%%%%%%%%%%%%%%%%%%%%%%%%%%%%%%%%%%%%%%%%%%%%%%%%%%%%%%%%%%%%%%%
%%%%%%%%%%%%%%%%%%%%%%%%%%%%%%%%%%% BIBLIOGRAPHY %%%%%%%%%%%%%%%%%%%%%%%%%%%%%%%%%%%%%%%
%%%%%%%%%%%%%%%%%%%%%%%%%%%%%%%%%%%%%%%%%%%%%%%%%%%%%%%%%%%%%%%%%%%%%%%%%%%%%%%%%%%%%%%%

% - use BibTeX with the regular commands:
   \bibliographystyle{aa} % style aa.bst
   \bibliography{3C273} % your references Yourfile.bib

\begin{thebibliography}{65}
\expandafter\ifx\csname natexlab\endcsname\relax\def\natexlab#1{#1}\fi

\bibitem[{{An} {et~al.}(2020){An}, {Mohan}, {Zhang}, {Frey}, {Yang},
  {Gab{\'a}nyi}, {Gurvits}, {Paragi}, {Perger}, \& {Zheng}}]{2020An_NatComm}
{An}, T., {Mohan}, P., {Zhang}, Y., {et~al.} 2020, Nature Communications, 11,
  143

\bibitem[{{Asada} \& {Nakamura}(2012)}]{Asada12}
{Asada}, K. \& {Nakamura}, M. 2012, \apjl, 745, L28

\bibitem[{{Astropy Collaboration} {et~al.}(2018){Astropy Collaboration},
  {Price-Whelan}, {Sip{\H{o}}cz}, {G{\"u}nther}, {Lim}, {Crawford}, {Conseil},
  {Shupe}, {Craig}, {Dencheva}, {Ginsburg}, {Vand erPlas}, {Bradley},
  {P{\'e}rez-Su{\'a}rez}, {de Val-Borro}, {Aldcroft}, {Cruz}, {Robitaille},
  {Tollerud}, {Ardelean}, {Babej}, {Bach}, {Bachetti}, {Bakanov}, {Bamford},
  {Barentsen}, {Barmby}, {Baumbach}, {Berry}, {Biscani}, {Boquien}, {Bostroem},
  {Bouma}, {Brammer}, {Bray}, {Breytenbach}, {Buddelmeijer}, {Burke},
  {Calderone}, {Cano Rodr{\'\i}guez}, {Cara}, {Cardoso}, {Cheedella}, {Copin},
  {Corrales}, {Crichton}, {D'Avella}, {Deil}, {Depagne}, {Dietrich}, {Donath},
  {Droettboom}, {Earl}, {Erben}, {Fabbro}, {Ferreira}, {Finethy}, {Fox},
  {Garrison}, {Gibbons}, {Goldstein}, {Gommers}, {Greco}, {Greenfield},
  {Groener}, {Grollier}, {Hagen}, {Hirst}, {Homeier}, {Horton}, {Hosseinzadeh},
  {Hu}, {Hunkeler}, {Ivezi{\'c}}, {Jain}, {Jenness}, {Kanarek}, {Kendrew},
  {Kern}, {Kerzendorf}, {Khvalko}, {King}, {Kirkby}, {Kulkarni}, {Kumar},
  {Lee}, {Lenz}, {Littlefair}, {Ma}, {Macleod}, {Mastropietro}, {McCully},
  {Montagnac}, {Morris}, {Mueller}, {Mumford}, {Muna}, {Murphy}, {Nelson},
  {Nguyen}, {Ninan}, {N{\"o}the}, {Ogaz}, {Oh}, {Parejko}, {Parley}, {Pascual},
  {Patil}, {Patil}, {Plunkett}, {Prochaska}, {Rastogi}, {Reddy Janga},
  {Sabater}, {Sakurikar}, {Seifert}, {Sherbert}, {Sherwood-Taylor}, {Shih},
  {Sick}, {Silbiger}, {Singanamalla}, {Singer}, {Sladen}, {Sooley},
  {Sornarajah}, {Streicher}, {Teuben}, {Thomas}, {Tremblay}, {Turner},
  {Terr{\'o}n}, {van Kerkwijk}, {de la Vega}, {Watkins}, {Weaver}, {Whitmore},
  {Woillez}, {Zabalza}, \& {Astropy Contributors}}]{Astropy:2018}
{Astropy Collaboration}, {Price-Whelan}, A.~M., {Sip{\H{o}}cz}, B.~M., {et~al.}
  2018, \aj, 156, 123

\bibitem[{{Astropy Collaboration} {et~al.}(2013){Astropy Collaboration},
  {Robitaille}, {Tollerud}, {Greenfield}, {Droettboom}, {Bray}, {Aldcroft},
  {Davis}, {Ginsburg}, {Price-Whelan}, {Kerzendorf}, {Conley}, {Crighton},
  {Barbary}, {Muna}, {Ferguson}, {Grollier}, {Parikh}, {Nair}, {Unther},
  {Deil}, {Woillez}, {Conseil}, {Kramer}, {Turner}, {Singer}, {Fox}, {Weaver},
  {Zabalza}, {Edwards}, {Azalee Bostroem}, {Burke}, {Casey}, {Crawford},
  {Dencheva}, {Ely}, {Jenness}, {Labrie}, {Lim}, {Pierfederici}, {Pontzen},
  {Ptak}, {Refsdal}, {Servillat}, \& {Streicher}}]{Astropy:2013}
{Astropy Collaboration}, {Robitaille}, T.~P., {Tollerud}, E.~J., {et~al.} 2013,
  \aap, 558, A33

\bibitem[{{Attridge} {et~al.}(1999){Attridge}, {Roberts}, \&
  {Wardle}}]{1999ApJ...518L..87A}
{Attridge}, J.~M., {Roberts}, D.~H., \& {Wardle}, J. F.~C. 1999, \apjl, 518,
  L87

\bibitem[{{Beuchert} {et~al.}(2018){Beuchert}, {Kadler}, {Perucho},
  {Gro{\ss}berger}, {Schulz}, {Agudo}, {Casadio}, {G{\'o}mez}, {Gurwell},
  {Homan}, {Kovalev}, {Lister}, {Markoff}, {Molina}, {Pushkarev}, {Ros},
  {Savolainen}, {Steinbring}, {Thum}, \& {Wilms}}]{2018A&A...610A..32B}
{Beuchert}, T., {Kadler}, M., {Perucho}, M., {et~al.} 2018, \aap, 610, A32

\bibitem[{{Blandford} \& {Payne}(1982)}]{1982MNRAS.199..883B}
{Blandford}, R.~D. \& {Payne}, D.~G. 1982, \mnras, 199, 883

\bibitem[{{Blandford} \& {Znajek}(1977)}]{1977MNRAS.179..433B}
{Blandford}, R.~D. \& {Znajek}, R.~L. 1977, \mnras, 179, 433

\bibitem[{{Boccardi} {et~al.}(2016){Boccardi}, {Krichbaum}, {Bach}, {Bremer},
  \& {Zensus}}]{2016A&A...588L...9B}
{Boccardi}, B., {Krichbaum}, T.~P., {Bach}, U., {Bremer}, M., \& {Zensus},
  J.~A. 2016, \aap, 588, L9

\bibitem[{{Bruni} {et~al.}(2016){Bruni}, {Anderson}, {Alef}, {Rottmann},
  {Lobanov}, \& {Zensus}}]{2016Galax...4...55B}
{Bruni}, G., {Anderson}, J., {Alef}, W., {et~al.} 2016, Galaxies, 4, 55

\bibitem[{{Bruni} {et~al.}(2017){Bruni}, {G{\'o}mez}, {Casadio}, {Lobanov},
  {Kovalev}, {Sokolovsky}, {Lisakov}, {Bach}, {Marscher}, {Jorstad},
  {Anderson}, {Krichbaum}, {Savolainen}, {Vega-Garc{\'\i}a}, {Fuentes},
  {Zensus}, {Alberdi}, {Lee}, {Lu}, {P{\'e}rez-Torres}, \&
  {Ros}}]{2017A&A...604A.111B}
{Bruni}, G., {G{\'o}mez}, J.~L., {Casadio}, C., {et~al.} 2017, \aap, 604, A111

\bibitem[{{Bruni} {et~al.}(2020){Bruni}, {Savolainen}, {G{\'o}mez}, {Lobanov},
  {Kovalev}, {RadioAstron AGN Imaging Team}, \& {KSP
  Team}}]{2020AdSpR..65..712B}
{Bruni}, G., {Savolainen}, T., {G{\'o}mez}, J.~L., {et~al.} 2020, Advances in
  Space Research, 65, 712

\bibitem[{{Celotti} {et~al.}(2001){Celotti}, {Ghisellini}, \&
  {Chiaberge}}]{2001MNRAS.321L...1C}
{Celotti}, A., {Ghisellini}, G., \& {Chiaberge}, M. 2001, \mnras, 321, L1

\bibitem[{{Cohen} {et~al.}(2015){Cohen}, {Meier}, {Arshakian}, {Clausen-Brown},
  {Homan}, {Hovatta}, {Kovalev}, {Lister}, {Pushkarev}, {Richards}, \&
  {Savolainen}}]{2015ApJ...803....3C}
{Cohen}, M.~H., {Meier}, D.~L., {Arshakian}, T.~G., {et~al.} 2015, \apj, 803, 3

\bibitem[{{D'Arcangelo} {et~al.}(2009){D'Arcangelo}, {Marscher}, {Jorstad},
  {Smith}, {Larionov}, {Hagen-Thorn}, {Williams}, {Gear}, {Clemens}, {Sarcia},
  {Grabau}, {Tollestrup}, {Buie}, {Taylor}, \& {Dunham}}]{2009ApJ...697..985D}
{D'Arcangelo}, F.~D., {Marscher}, A.~P., {Jorstad}, S.~G., {et~al.} 2009, \apj,
  697, 985

\bibitem[{{Davis} \& {Tchekhovskoy}(2020)}]{2020ARA&A..58..407D}
{Davis}, S.~W. \& {Tchekhovskoy}, A. 2020, \araa, 58, 407

\bibitem[{{Ghisellini} {et~al.}(2005){Ghisellini}, {Tavecchio}, \&
  {Chiaberge}}]{2005A&A...432..401G}
{Ghisellini}, G., {Tavecchio}, F., \& {Chiaberge}, M. 2005, \aap, 432, 401

\bibitem[{{Giovannini} {et~al.}(2018){Giovannini}, {Savolainen}, {Orienti},
  {Nakamura}, {Nagai}, {Kino}, {Giroletti}, {Hada}, {Bruni}, {Kovalev},
  {Anderson}, {D'Ammando}, {Hodgson}, {Honma}, {Krichbaum}, {Lee}, {Lico},
  {Lisakov}, {Lobanov}, {Petrov}, {Sohn}, {Sokolovsky}, {Voitsik}, {Zensus}, \&
  {Tingay}}]{2018NatAs...2..472G}
{Giovannini}, G., {Savolainen}, T., {Orienti}, M., {et~al.} 2018, Nature
  Astronomy, 2, 472

\bibitem[{{Giroletti} {et~al.}(2004){Giroletti}, {Giovannini}, {Feretti},
  {Cotton}, {Edwards}, {Lara}, {Marscher}, {Mattox}, {Piner}, \&
  {Venturi}}]{2004ApJ...600..127G}
{Giroletti}, M., {Giovannini}, G., {Feretti}, L., {et~al.} 2004, \apj, 600, 127

\bibitem[{{G{\'o}mez} {et~al.}(2016){G{\'o}mez}, {Lobanov}, {Bruni}, {Kovalev},
  {Marscher}, {Jorstad}, {Mizuno}, {Bach}, {Sokolovsky}, {Anderson}, {Galindo},
  {Kardashev}, \& {Lisakov}}]{2016ApJ...817...96G}
{G{\'o}mez}, J.~L., {Lobanov}, A.~P., {Bruni}, G., {et~al.} 2016, \apj, 817, 96

\bibitem[{{Gurvits} {et~al.}(1997){Gurvits}, {Schilizzi}, {Miley}, {Peck},
  {Bremer}, {Roettgering}, \& {van Breugel}}]{1997A&A...318...11G}
{Gurvits}, L.~I., {Schilizzi}, R.~T., {Miley}, G.~K., {et~al.} 1997, \aap, 318,
  11

\bibitem[{{Hardee}(2007)}]{2007ApJ...664...26H}
{Hardee}, P.~E. 2007, \apj, 664, 26

\bibitem[{{Hazard} {et~al.}(1963){Hazard}, {Mackey}, \&
  {Shimmins}}]{1963Natur.197.1037H}
{Hazard}, C., {Mackey}, M.~B., \& {Shimmins}, A.~J. 1963, \nat, 197, 1037

\bibitem[{Hunter(2007)}]{Hunter:2007}
Hunter, J.~D. 2007, Computing In Science \& Engineering, 9, 90

\bibitem[{{Johnson} {et~al.}(2016){Johnson}, {Kovalev}, {Gwinn}, {Gurvits},
  {Narayan}, {Macquart}, {Jauncey}, {Voitsik}, {Anderson}, {Sokolovsky}, \&
  {Lisakov}}]{2016ApJ...820L..10J}
{Johnson}, M.~D., {Kovalev}, Y.~Y., {Gwinn}, C.~R., {et~al.} 2016, \apjl, 820,
  L10

\bibitem[{{Jorstad} {et~al.}(2005){Jorstad}, {Marscher}, {Lister}, {Stirling},
  {Cawthorne}, {Gear}, {G{\'o}mez}, {Stevens}, {Smith}, {Forster}, \&
  {Robson}}]{2005AJ....130.1418J}
{Jorstad}, S.~G., {Marscher}, A.~P., {Lister}, M.~L., {et~al.} 2005, \aj, 130,
  1418

\bibitem[{{Kardashev} {et~al.}(2013){Kardashev}, {Khartov}, {Abramov},
  {Avdeev}, {Alakoz}, {Aleksandrov}, {Ananthakrishnan}, {Andreyanov},
  {Andrianov}, {Antonov}, {Artyukhov}, {Arkhipov}, {Baan}, {Babakin},
  {Babyshkin}, {Bartel'}, {Belousov}, {Belyaev}, {Berulis}, {Burke},
  {Biryukov}, {Bubnov}, {Burgin}, {Busca}, {Bykadorov}, {Bychkova},
  {Vasil'kov}, {Wellington}, {Vinogradov}, {Wietfeldt}, {Voitsik},
  {Gvamichava}, {Girin}, {Gurvits}, {Dagkesamanskii}, {D'Addario},
  {Giovannini}, {Jauncey}, {Dewdney}, {D'yakov}, {Zharov}, {Zhuravlev},
  {Zaslavskii}, {Zakhvatkin}, {Zinov'ev}, {Ilinen}, {Ipatov}, {Kanevskii},
  {Knorin}, {Casse}, {Kellermann}, {Kovalev}, {Kovalev}, {Kovalenko}, {Kogan},
  {Komaev}, {Konovalenko}, {Kopelyanskii}, {Korneev}, {Kostenko}, {Kotik},
  {Kreisman}, {Kukushkin}, {Kulishenko}, {Cooper}, {Kut'kin}, {Cannon},
  {Larionov}, {Lisakov}, {Litvinenko}, {Likhachev}, {Likhacheva}, {Lobanov},
  {Logvinenko}, {Langston}, {McCracken}, {Medvedev}, {Melekhin}, {Menderov},
  {Murphy}, {Mizyakina}, {Mozgovoi}, {Nikolaev}, {Novikov}, {Novikov},
  {Oreshko}, {Pavlenko}, {Pashchenko}, {Ponomarev}, {Popov}, {Pravin-Kumar},
  {Preston}, {Pyshnov}, {Rakhimov}, {Rozhkov}, {Romney}, {Rocha}, {Rudakov},
  {R{\"a}is{\"a}nen}, {Sazankov}, {Sakharov}, {Semenov}, {Serebrennikov},
  {Schilizzi}, {Skulachev}, {Slysh}, {Smirnov}, {Smith}, {Soglasnov},
  {Sokolovskii}, {Sondaar}, {Stepan'yants}, {Turygin}, {Turygin}, {Tuchin},
  {Urpo}, {Fedorchuk}, {Finkel'shtein}, {Fomalont}, {Fejes}, {Fomina},
  {Khapin}, {Tsarevskii}, {Zensus}, {Chuprikov}, {Shatskaya}, {Shapirovskaya},
  {Sheikhet}, {Shirshakov}, {Schmidt}, {Shnyreva}, {Shpilevskii}, {Ekers}, \&
  {Yakimov}}]{2013ARep...57..153K}
{Kardashev}, N.~S., {Khartov}, V.~V., {Abramov}, V.~V., {et~al.} 2013,
  Astronomy Reports, 57, 153

\bibitem[{{Kim} {et~al.}(2018){Kim}, {Krichbaum}, {Lu}, {Ros}, {Bach},
  {Bremer}, {de Vicente}, {Lindqvist}, \& {Zensus}}]{2018A&A...616A.188K}
{Kim}, J.~Y., {Krichbaum}, T.~P., {Lu}, R.~S., {et~al.} 2018, \aap, 616, A188

\bibitem[{{Komissarov} \& {Porth}(2021)}]{2021NewAR..9201610K}
{Komissarov}, S. \& {Porth}, O. 2021, \nar, 92, 101610

\bibitem[{{Kovalev} {et~al.}(2016){Kovalev}, {Kardashev}, {Kellermann},
  {Lobanov}, {Johnson}, {Gurvits}, {Voitsik}, {Zensus}, {Anderson}, {Bach},
  {Jauncey}, {Ghigo}, {Ghosh}, {Kraus}, {Kovalev}, {Lisakov}, {Petrov},
  {Romney}, {Salter}, \& {Sokolovsky}}]{2016ApJ...820L...9K}
{Kovalev}, Y.~Y., {Kardashev}, N.~S., {Kellermann}, K.~I., {et~al.} 2016,
  \apjl, 820, L9

\bibitem[{{Kovalev} {et~al.}(2020{\natexlab{a}}){Kovalev}, {Kardashev},
  {Sokolovsky}, {Voitsik}, {An}, {Anderson}, {Andrianov}, {Avdeev}, {Bartel},
  {Bignall}, {Burgin}, {Edwards}, {Ellingsen}, {Frey}, {Garc{\'\i}a-Mir{\'o}},
  {Gawro{\'n}ski}, {Ghigo}, {Ghosh}, {Giovannini}, {Girin}, {Giroletti},
  {Gurvits}, {Jauncey}, {Horiuchi}, {Ivanov}, {Kharinov}, {Koay}, {Kostenko},
  {Kovalenko}, {Kovalev}, {Kravchenko}, {Kunert-Bajraszewska}, {Kutkin},
  {Likhachev}, {Lisakov}, {Litovchenko}, {McCallum}, {Melis}, {Melnikov},
  {Migoni}, {Nair}, {Pashchenko}, {Phillips}, {Polatidis}, {Pushkarev},
  {Quick}, {Rakhimov}, {Reynolds}, {Rizzo}, {Rudnitskiy}, {Savolainen},
  {Shakhvorostova}, {Shatskaya}, {Shen}, {Shchurov}, {Vermeulen}, {de Vicente},
  {Wolak}, {Zensus}, \& {Zuga}}]{2020AdSpR..65..705K}
{Kovalev}, Y.~Y., {Kardashev}, N.~S., {Sokolovsky}, K.~V., {et~al.}
  2020{\natexlab{a}}, Advances in Space Research, 65, 705

\bibitem[{{Kovalev} {et~al.}(2020{\natexlab{b}}){Kovalev}, {Pushkarev},
  {Nokhrina}, {Plavin}, {Beskin}, {Chernoglazov}, {Lister}, \&
  {Savolainen}}]{2020MNRAS.495.3576K}
{Kovalev}, Y.~Y., {Pushkarev}, A.~B., {Nokhrina}, E.~E., {et~al.}
  2020{\natexlab{b}}, \mnras, 495, 3576

\bibitem[{{Lisakov} {et~al.}(2017){Lisakov}, {Kovalev}, {Savolainen},
  {Hovatta}, \& {Kutkin}}]{2017MNRAS.468.4478L}
{Lisakov}, M.~M., {Kovalev}, Y.~Y., {Savolainen}, T., {Hovatta}, T., \&
  {Kutkin}, A.~M. 2017, \mnras, 468, 4478

\bibitem[{{Lister} {et~al.}(2019){Lister}, {Homan}, {Hovatta}, {Kellermann},
  {Kiehlmann}, {Kovalev}, {Max-Moerbeck}, {Pushkarev}, {Readhead}, {Ros}, \&
  {Savolainen}}]{2019ApJ...874...43L}
{Lister}, M.~L., {Homan}, D.~C., {Hovatta}, T., {et~al.} 2019, \apj, 874, 43

\bibitem[{{Lobanov}(1998)}]{1998A&AS..132..261L}
{Lobanov}, A.~P. 1998, \aaps, 132, 261

\bibitem[{{Lobanov} {et~al.}(1997){Lobanov}, {Carrara}, \&
  {Zensus}}]{1997VA.....41..253L}
{Lobanov}, A.~P., {Carrara}, E., \& {Zensus}, J.~A. 1997, Vistas in Astronomy,
  41, 253

\bibitem[{{Lobanov} \& {Zensus}(1999)}]{1999ApJ...521..509L}
{Lobanov}, A.~P. \& {Zensus}, J.~A. 1999, \apj, 521, 509

\bibitem[{{Lobanov} \& {Zensus}(2001)}]{2001Sci...294..128L}
{Lobanov}, A.~P. \& {Zensus}, J.~A. 2001, Science, 294, 128

\bibitem[{{Lobanov} {et~al.}(2000){Lobanov}, {Zensus}, {Abraham}, {Carrara},
  {Unwin}, {Hirabayashi}, \& {Bushimata}}]{2000AdSpR..26..669L}
{Lobanov}, A.~P., {Zensus}, J.~A., {Abraham}, Z., {et~al.} 2000, Advances in
  Space Research, 26, 669

\bibitem[{{Marscher} \& {Gear}(1985)}]{1985ApJ...298..114M}
{Marscher}, A.~P. \& {Gear}, W.~K. 1985, \apj, 298, 114

\bibitem[{{Mart{\'\i}} {et~al.}(2016){Mart{\'\i}}, {Perucho}, \&
  {G{\'o}mez}}]{2016ApJ...831..163M}
{Mart{\'\i}}, J.~M., {Perucho}, M., \& {G{\'o}mez}, J.~L. 2016, \apj, 831, 163

\bibitem[{McKinney(2010)}]{mckinneyprocscipy2010}
McKinney, W. 2010, in Proceedings of the 9th Python in Science Conference, ed.
  S.~van~der Walt \& J.~Millman, 51 -- 56

\bibitem[{{Mertens} {et~al.}(2016){Mertens}, {Lobanov}, {Walker}, \&
  {Hardee}}]{2016A&A...595A..54M}
{Mertens}, F., {Lobanov}, A.~P., {Walker}, R.~C., \& {Hardee}, P.~E. 2016,
  \aap, 595, A54

\bibitem[{{Mimica} {et~al.}(2015){Mimica}, {Giannios}, {Metzger}, \&
  {Aloy}}]{2015MNRAS.450.2824M}
{Mimica}, P., {Giannios}, D., {Metzger}, B.~D., \& {Aloy}, M.~A. 2015, \mnras,
  450, 2824

\bibitem[{{Nokhrina} {et~al.}(2020){Nokhrina}, {Kovalev}, \&
  {Pushkarev}}]{2020MNRAS.498.2532N}
{Nokhrina}, E.~E., {Kovalev}, Y.~Y., \& {Pushkarev}, A.~B. 2020, \mnras, 498,
  2532

\bibitem[{{Ogihara} {et~al.}(2019){Ogihara}, {Takahashi}, \&
  {Toma}}]{2019ApJ...877...19O}
{Ogihara}, T., {Takahashi}, K., \& {Toma}, K. 2019, \apj, 877, 19

\bibitem[{{Oke}(1963)}]{1963Natur.197.1040O}
{Oke}, J.~B. 1963, \nat, 197, 1040

\bibitem[{{Pelletier} \& {Roland}(1989)}]{1989A&A...224...24P}
{Pelletier}, G. \& {Roland}, J. 1989, \aap, 224, 24

\bibitem[{{Perucho} {et~al.}(2012){Perucho}, {Kovalev}, {Lobanov}, {Hardee}, \&
  {Agudo}}]{2012ApJ...749...55P}
{Perucho}, M., {Kovalev}, Y.~Y., {Lobanov}, A.~P., {Hardee}, P.~E., \& {Agudo},
  I. 2012, \apj, 749, 55

\bibitem[{{Perucho} {et~al.}(2006){Perucho}, {Lobanov}, {Mart{\'\i}}, \&
  {Hardee}}]{2006A&A...456..493P}
{Perucho}, M., {Lobanov}, A.~P., {Mart{\'\i}}, J.~M., \& {Hardee}, P.~E. 2006,
  \aap, 456, 493

\bibitem[{{Plavin} {et~al.}(2019){Plavin}, {Kovalev}, {Pushkarev}, \&
  {Lobanov}}]{2019MNRAS.485.1822P}
{Plavin}, A.~V., {Kovalev}, Y.~Y., {Pushkarev}, A.~B., \& {Lobanov}, A.~P.
  2019, \mnras, 485, 1822

\bibitem[{{Pushkarev} {et~al.}(2005){Pushkarev}, {Gabuzda}, {Vetukhnovskaya},
  \& {Yakimov}}]{2005MNRAS.356..859P}
{Pushkarev}, A.~B., {Gabuzda}, D.~C., {Vetukhnovskaya}, Y.~N., \& {Yakimov},
  V.~E. 2005, \mnras, 356, 859

\bibitem[{{Pushkarev} {et~al.}(2017){Pushkarev}, {Kovalev}, {Lister}, \&
  {Savolainen}}]{2017MNRAS.468.4992P}
{Pushkarev}, A.~B., {Kovalev}, Y.~Y., {Lister}, M.~L., \& {Savolainen}, T.
  2017, \mnras, 468, 4992

\bibitem[{{Savolainen} {et~al.}(2006){Savolainen}, {Wiik}, {Valtaoja}, \&
  {Tornikoski}}]{2006A&A...446...71S}
{Savolainen}, T., {Wiik}, K., {Valtaoja}, E., \& {Tornikoski}, M. 2006, \aap,
  446, 71

\bibitem[{{Schmidt}(1963)}]{1963Natur.197.1040S}
{Schmidt}, M. 1963, \nat, 197, 1040

\bibitem[{{Schulz} {et~al.}(2020){Schulz}, {Kadler}, {Ros}, {Perucho},
  {Krichbaum}, {Agudo}, {Beuchert}, {Lindqvist}, {Mannheim}, {Wilms}, \&
  {Zensus}}]{2020A&A...644A..85S}
{Schulz}, R., {Kadler}, M., {Ros}, E., {et~al.} 2020, \aap, 644, A85

\bibitem[{{Shepherd}(1997)}]{1997ASPC..125...77S}
{Shepherd}, M.~C. 1997, in Astronomical Society of the Pacific Conference
  Series, Vol. 125, Astronomical Data Analysis Software and Systems VI, ed.
  G.~{Hunt} \& H.~{Payne}, 77

\bibitem[{{Sikora} {et~al.}(2007){Sikora}, {Stawarz}, \&
  {Lasota}}]{2007ApJ...658..815S}
{Sikora}, M., {Stawarz}, {\L}., \& {Lasota}, J.-P. 2007, \apj, 658, 815

\bibitem[{{Sol} {et~al.}(1989){Sol}, {Pelletier}, \&
  {Asseo}}]{1989MNRAS.237..411S}
{Sol}, H., {Pelletier}, G., \& {Asseo}, E. 1989, \mnras, 237, 411

\bibitem[{{Tavecchio} \& {Ghisellini}(2008)}]{2008MNRAS.386..945T}
{Tavecchio}, F. \& {Ghisellini}, G. 2008, \mnras, 386, 945

\bibitem[{{van der Walt} {et~al.}(2011){van der Walt}, {Colbert}, \&
  {Varoquaux}}]{5725236}
{van der Walt}, S., {Colbert}, S.~C., \& {Varoquaux}, G. 2011, Computing in
  Science Engineering, 13, 22

\bibitem[{{Vega-Garc{\'i}a}(2018)}]{kups9379}
{Vega-Garc{\'i}a}, L. 2018, PhD thesis, Universit{\"a}t zu K{\"o}ln,
  https://kups.ub.uni-koeln.de/9379/

\bibitem[{{Vega-Garc{\'\i}a} {et~al.}(2019){Vega-Garc{\'\i}a}, {Perucho}, \&
  {Lobanov}}]{2019A&A...627A..79V}
{Vega-Garc{\'\i}a}, L., {Perucho}, M., \& {Lobanov}, A.~P. 2019, \aap, 627, A79

\bibitem[{{Walker} {et~al.}(2018){Walker}, {Hardee}, {Davies}, {Ly}, \&
  {Junor}}]{2018ApJ...855..128W}
{Walker}, R.~C., {Hardee}, P.~E., {Davies}, F.~B., {Ly}, C., \& {Junor}, W.
  2018, \apj, 855, 128

\bibitem[{{Xie} {et~al.}(2012){Xie}, {Lei}, {Zou}, {Wang}, {Wu}, \&
  {Wang}}]{2012RAA....12..817X}
{Xie}, W., {Lei}, W.-H., {Zou}, Y.-C., {et~al.} 2012, Research in Astronomy and
  Astrophysics, 12, 817

\end{thebibliography}

%%%%%%%%%%%%%%%%%%%%%%%%%%%%%%%%%%%%%%%%%%%%%%%%%%%%%%%%%%%%%%%%%%%%%%%%%%%%%%%%%%%%%%%%
%%%%%%%%%%%%%%%%%%%%%%%%%%%%%%%%%%%%%%%%%%%%%%%%%%%%%%%%%%%%%%%%%%%%%%%%%%%%%%%%%%%%%%%%
\end{document}